\documentclass[amssymb,11pt]{article}
\usepackage{latex8}
\usepackage{times}

\usepackage{amsmath}
\usepackage{mathbbold}
\usepackage{amssymb}
\usepackage{graphicx}

\def\>{\ensuremath{\rangle}}
\def\<{\ensuremath{\langle}}

\newtheorem{thm}{Theorem}[section]

\newtheorem{lem}{Lemma}[section]
\newtheorem{defn}{Definition}[section]
\newtheorem{prop}{Proposition}[section]
\newtheorem{exam}{Example}[section]

\begin{document}

\title{Verification of Quantum Programs}

\author{Mingsheng Ying, Nengkun Yu, Yuan Feng, and Runyao Duan\\
           QCIS, FEIT, University of Technology, Sydney, and\\ TNList, Dept. of CS, Tsinghua
University\\
          mying@it.uts.edu.au}

\maketitle

\begin{abstract} This paper develops verification methodology for
quantum programs, and the contribution of the paper is
two-fold:\begin{itemize}
\item Sharir, Pnueli and Hart \textit{[SIAM J. Comput.
13(1984)292-314]} presented a general method for proving properties
of probabilistic programs, in which a probabilistic program is
modeled by a Markov chain and an assertion on the output
distribution is extended into an invariant assertion on all
intermediate distributions. Their method is essentially a
probabilistic generalization of the classical Floyd inductive
assertion method. In this paper, we consider quantum programs
modeled by quantum Markov chains which are defined by
super-operators. It is shown that the Sharir-Pnueli-Hart method can
be elegantly generalized to quantum programs by exploiting the
Schr\"odinger-Heisenberg duality between quantum states and
observables. In particular, a completeness theorem for the
Sharir-Pnueli-Hart verification method of quantum programs is
established. \item As indicated by the completeness theorem, the
Sharir-Pnueli-Hart method is in principle effective for verifying
all properties of quantum programs that can be expressed in terms of
Hermitian operators (observables). But it is not feasible for many
practical applications because of the complicated calculation
involved in the verification. For the case of finite-dimensional
state spaces, we find a method for verification of quantum programs
much simpler than the Sharir-Pnueli-Hart method by employing the
matrix representation of super-operators and Jordan decomposition of
matrices. In particular, this method enables us to compute easily
the average running time and even to analyze some interesting
long-run behaviors of quantum programs in a finite-dimensional state
space.
\end{itemize}\end{abstract}

\section{Introduction}

The need of techniques for verification of quantum programs comes from two areas:\begin{itemize}
\item Since the early 1980s, various quantum communication protocols have been proposed, and now
quantum cryptographic systems are already commercially available from Id Quantique, MagiQ
Technologies, SmartQuantum and NEC. A salient advantage of quantum communication over classical
communication is that its security is provable based on the principles of quantum mechanics.
However, it is very difficult to guarantee the correctness of protocols at the stage of design.
Some simple classical protocols were finally found to have fundamental flaws. The case of quantum protocols
is even worse because of the combined weirdness of cryptographic protocols and the quantum world.
Indeed, an attack on Id Quantique's commercial quantum cryptographic system was recently reported~\cite{XQL10}.
Since quantum communication protocols can be expressed as quantum programs, the need in this area is urgent.
\item Steady progress in the techniques of quantum devices has made people widely believe that large-scalable
and functional quantum
computers will be eventually built. Once quantum computers come into truth, quantum programming
techniques will play a key role in exploiting the power of quantum computers. It is likely that programmers
will commit faults more often in designing programs for quantum computers than programming classical computers,
since human intuition is much better adapted to the classical world than the quantum world. Therefore,
verification techniques and tools will be indispensable to warrant correctness of quantum programs.
The need in this area is not urgent because quantum hardware is still in its infancy, but it will be huge
when quantum computers are commercialized.
\end{itemize}

Several techniques for verifying quantum programs have already been
developed in the last 10 years~\cite{G06,Se04a}. Various formal
semantics have been defined for the existing quantum programming
languages, and they can be used to reason about the correctness of
quantum programs written in these languages; for example, an
operational semantics was given to Sanders and Zuliani's language
qGCL~\cite{SZ00,Zu01}, a denotational semantics was defined for
Selinger's QPL~\cite{Se04} by representing quantum programs as
super-operators, and a denotational semantics of Altenkirch and
Grattage's language QML~\cite{AG05} was described in
category-theoretic terms. Another research line is
language-independent approach; for example, D'Hondt and
Panangaden~\cite{DP06} proposed the Hermitian operator
representation of quantum predicates and thus introduced an
intrinsic notion of quantum weakest preconditions, and quantum
predicate transformer semantics was further developed by Ying et
al.~\cite{YCFD07,YDFJ10}. Also, a few program logics have been
introduced into the quantum setting; for example, Baltag and
Smets~\cite{BS06} proposed a dynamic logic formalism of information
flows in quantum systems, Brunet and Jorrand~\cite{BJ04} found a way
of applying Birkhoff and von Neumann's quantum logic to reason about
quantum programs, Chadha, Mateus and Sernadas~\cite{CMS06} presented
a Floyd-Hoare logic for quantum programs with bounded iterations,
and Feng et al.~\cite{FDJY07} introduced some useful proof rules for
reasoning about quantum loops.

Quantum mechanics is an intrinsically probabilistic theory. So, it
is natural to see whether the successful methods for verifying
probabilistic programs can be generalized to the quantum case.
Sharir, Pnueli and Hart~\cite{SPH84} introduced a general method for
verifying probabilistic programs. The method is
language-independent, and it models a probabilistic program as a
Markov chain defined on the program states. An initial distribution
of the program state is given, and a set of terminal states is
assumed. At each step of the execution, the distribution of the next
state is determined by the distribution of the current state and the
matrix of transition probabilities of the Markov chain, and the
terminal states remain unchanged. The Markov chain model is very
popular in the current studies of probabilistic programs~\cite{CY95,
dA98, BEK05}. The property of the program to be verified is an
assertion on the distribution of the terminal states, and it is
represented by a linear functional of the state distribution. The
key idea of this method is to extend it into an invariant assertion
on all intermediate distributions of the program states, including
the initial one. Thus, the problem of verifying the program is
reduced to checking whether the initial distribution satisfies the
invariant assertion without computing the distribution of the
terminal states, which is usually formidable. This method is
essentially a probabilistic generalization of the classical Floyd
inductive assertion method for deterministic programs. The
Sharir-Pnueli-Hart method can be used to infer statistical
properties of a deterministic program, such as its average running
time, the expected value of certain output variable, the probability
of program termination, etc., when the inputs of the programs are
drawn with a known probability distribution. Also, it can be used to
analyze random algorithms in which the decision at nondeterministic
branching is made according to certain known probability
distribution.

\textbf{Our Contribution:} The first aim of this paper is to develop
a verification method for quantum programs in the Sharir-Pnueli-Hart
style. According to the Hilbert space formalism of quantum
mechanics, the state space of a quantum program is supposed to be a
separable Hilbert space. A state of the program is described by a
density operator. Then program is modeled by the quantum counterpart
of a Markov chain, namely, a quantum Markov chain. The transitions
in the quantum Markov chain is depicted by a super-operator, which
transforms density operators to themselves. The reasonableness of
this model can be argued from the following three aspects:
\begin{itemize}\item First, super-operators are one of the most popular mathematical
formalisms of physically realizable operations allowed by quantum
mechanics~\cite{NC00}, including of course all the operations that
quantum computers can perform. \item Second, this model coincides
with Selinger's insight of representing quantum programs as
super-operators~\cite{Se04}. \item Third, super-operators are widely
used as the mathematical model of quantum communication channels so
that the technique developed in this paper can be conveniently
applied in verifying quantum communication protocols.\end{itemize}

By exploiting the Schr\"odinger-Heisenberg duality between quantum
states and observables, we are able to generalize the
Sharir-Pnueli-Hart method to reason about quantum programs within
the quantum Markov chain framework. In particular, a completeness
theorem for the Sharir-Pnueli-Hart verification method of quantum
programs is established, which indicates that the method is in
principle effective for verifying all properties of quantum programs
that can be expressed in terms of Hermitian operators (observables).

Although the completeness theorem guarantees its universal
effectiveness, the quantum Sharir-Pnueli-Hart method is often not
feasible in practice since usually a very complicated calculation
involving iterations of super-operators is required in its
applications. For the case of finite-dimensional state spaces, each
super-operator enjoys a matrix representation which is easier to
manipulate than the super-operator itself. By using this powerful
mathematical tool, we find a method for verification of quantum
programs in finite-dimensional state spaces much simpler than the
Sharir-Pnueli-Hart method. The effectiveness of this simplified
verification method depends upon the condition that the modules of
all eigenvalues of a certain matrix are strictly smaller than $1$.
Fortunately, a delicate matrix analysis allows us to vanish the
Jordan blocks of the matrix corresponding to those eigenvalues with
module $1$, and thus warrants the effectiveness of the simplified
method.

Another difficulty in the practical applications of the
Sharir-Pnueli-Hart method often arises in checking termination
condition. We are able to provide some alternate termination
conditions much easier to check than the original one in the case of
finite-dimensional state spaces also by employing the matrix
representation of super-operators and Jordan decomposition of
matrices.

The paper is organized as follows: Some preliminaries are presented
in Sec.~\ref{prel}. In Sec.~\ref{main}, the Quantum
Sharir-Pnueli-Hart method is introduced, and a simple example is
provided to illustrate how to use this method . Also, correctness
and completeness of the method are proved. A simplified verification
method only suited to quantum programs in finite-dimensional state
spaces is proposed in Sec.~\ref{finite1} Sec.~\ref{finite2} is
devoted to examine termination of quantum programs in
finite-dimensional state spaces. For readability, the proofs of two
technical lemmas used in Sec.~\ref{finite1} and~\ref{finite2} are
postponed to the appendix.

\section{Preliminaries}\label{prel}

\subsection{Sharir-Pnueli-Hart Method}\label{SPH-p}

For convenience of the reader, we briefly review the Sharir-Pnueli-Hart method. A comparison of this method and
its quantum generalization presented in Sec.~\ref{main} is very helpful for the understanding of the latter.

Let $S$ be the set of program states, which is finite or countably
infinite. We assume that $T\subseteq S$ is the set of terminal
(absorbing) states, and write $I=S\setminus T$. A program is
understood as a probabilistic transition on $S$. A single step of
the program is represented by a matrix $P=(P_{ij})_{i,j\in S}$ of
transition probabilities, where $P_{ij}$ is the probability of
moving from state $i$ to state $j$ for each $i,j\in S$. It is
reasonable to assume that $P_{ij}=\delta_{ij}$ for all $j\in S$ if
$i\in T$ is a terminal state. We also assume an initial distribution
$\mu^{0}=(\mu_i^{0})_{i\in S}$ of the program state, where
$\mu_i^{0}$ is the probability that the program is initially in
state $i$ for each $i\in S$. So, a program can be modeled by a
Markov chain with transition probability matrix $P$ and initial
distribution $\mu^{0}$. An execution of the program is then a path
in the Markov chain; that is, a sequence of states in $S$. For any
$n\geq 0$, we define $\mu^{(n)}$ to be the distribution of the
program state after $n$ steps; that is, for each $i\in S$,
\begin{equation*}\begin{split} \mu_i^{(n)}&=Pr\{{\rm the\ program\
is\ in\ state}\ i\ {\rm after}\ n\ {\rm steps}\}\\ &=\sum_{j\in
S}\mu_j^{0}P_{ji}^{n},
\end{split}\end{equation*} where $P^{n}=(P^{n}_{ij})_{i,j\in S}$ is the $n$th power of $P$.
For any $i,j\in S$ and $n\geq 1$, we also define:

\begin{equation*}\begin{split}
f_{ij}^{(n)}&=Pr\{{\rm the\ program\ reaches\ state}\ j\ {\rm from\ state}\ i\\ &\ \ \ \ \ \ \ \ \ \ \ \ \ \ \ \ \ \ \ \ \ \ \ \ \ {\rm for\ the\ first\ time\ in\ exactly}\ n\ {\rm steps}\}\\
&=\sum_{i_1,...,i_{n-1}\in S\setminus\{j\}}\prod_{k=0}^{n-1}P_{i_ki_{k+1}},
\end{split}\end{equation*}where $i_0=i$ and $i_n=j$. We write: $$f_{ij}^{\ast}=\sum_{n=1}^{\infty}f_{ij}^{(n)}$$
for each $i,j\in S$. Intuitively, $f_{ij}^{\ast}$ is the probability
of ever getting to state $j$ from state $i$. Furthermore, for each
terminal state $j\in T$, we define $\mu_j^{\ast}$ to be the
probability that the program ever gets to state $j$:
$$\mu_j^{\ast}=\sum_{i\in S}\mu_i^{0}f_{ij}^{\ast},$$ and it is
obvious that $\mu^{\ast}=(\mu_j^{\ast})_{j\in T}$ is the (partial)
distribution of the program's terminal state. Thus, the program can
be seen as a distribution transformer, which sends the initial
distribution $\mu^{0}$ to the distribution $\mu^{\ast}$ of terminal
states, and this transformer can be considered as the denotational
(input-output) semantics of the program.

As one can imagine, in the general case, the distribution
$\mu^{\ast}$ of the program's terminal state is very difficult to
compute explicitly. However, Sharir, Pnueli and Hart~\cite{SPH84}
discovered an effective method to verify a class of properties of
the program. They first observed that a large class of important
properties of probabilistic programs, e.g. the expected running
time, the probability of termination, the expected value of some
program variable, can be represented by a linear functional of
$\mu^{\ast}$ of the form:
$$\psi (\mu^{\ast})=\sum_{j\in T}\mu_j^{\ast}\beta_j,$$ where $\beta_j\geq 0$ for all $j\in T$.
Their method for computing $\psi (\mu^{\ast})$ is indeed a
probabilistic generalization of the Floyd inductive assertion
method. The main idea is to find an invariant of the distribution
transformer so that one can trace back to the initial distribution
$\mu^{0}$ through all immediate distributions. More precisely, what
we need to do is to find a completion
$\overline{\beta}=(\overline{\beta}_i)_{i\in S}$ of the partial
vector $(\beta_j)_{j\in T}$ on the whole set of program states, with
$\overline{\beta}_j=\beta_j$ for all $j\in T$ and $\beta_i\geq 0$
for all $i\in I$. If $\overline{\beta}$ satisfies the following
conditions: \begin{itemize}
\item
(V1) $\varphi(\mu^{0})\stackrel{def}{=}\sum_{i\in
S}\mu_i^{0}\overline{\beta}_i<\infty$;
\item
(V2) $P\overline{\beta}=\overline{\beta}$;
\item
(V3) ($\overline{\beta}-$termination) $\sum_{i\in I}\mu_i^{(n)}\overline{\beta}_i\rightarrow 0$ when $n\rightarrow\infty$,
\end{itemize} then we conclude: \begin{itemize}\item (C) $\psi(\mu^{\ast})=\varphi(\mu^{0}).$\end{itemize}
Thus, computing the functional $\psi(\mu^{\ast})$ of the terminal
distribution $\mu^{\ast}$ can be done by computing a linear
functional $\varphi(\mu^{0})$ of the initial distribution $\mu^{0}$.

This method can be understood better if we compare it with the Floyd
method in the following way: Condition (V1) is a probabilistic
version of saying that $\varphi$ is true in the initial state; (V2)
means that $\overline{\beta}$ is an invariant of the program (a
fixed point of $P$), and thus it is analogous to the local
verification condition; (V3) means that the program terminates. The
implication from (V1), (V2) and (V3) to (C) is thus the analogue of
partial correctness.

The Sharir-Pnueli-Hart method was shown to be complete in the sense
that any partial vector $(\beta_j)_{j\in T}$ can be extended to
$\overline{\beta}$ required by the method:

\begin{thm} (Sharir, Pnueli and Hart~\cite{SPH84}) There always exists a completion
$\overline{\beta}=(\overline{\beta}_i)_{i\in S}$ of $(\beta_j)_{j\in
T}$, which satisfies (V2) and either satisfies (V1), (V3) and thus
(C), or else $\psi(\mu^{\ast})=\infty$. More explicitly, the least
fixed-point $\overline{\beta}$ of equation $P\beta=\beta$,
$\beta\geq 0$ is such a completion, with
$$\overline{\beta}_i=\sum_{j\in T}\beta_jf_{ij}^{\ast}$$ for every
$i\in I$.
\end{thm}

\subsection{Quantum Domains}

We assume that the reader is familiar with the Hilbert space
formalism of quantum theory. Let $H$ be a separable Hilbert space,
which is the state space of the quantum systems considered in this
paper. We adopt the Dirac notation, using $|\varphi\rangle,
|\psi\rangle$, ... to denote vectors in $H$. The inner product of
vectors $|\varphi\rangle, |\psi\rangle$ is written
$\langle\varphi|\psi\rangle$. The L\"{o}wner order of operators is
defined as follows: $A\sqsubseteq B$ if and only if $B-A$ is a
positive operator. Recall that a partial density operator is a
positive operator $\rho$ with its trace $tr(\rho)\leq 1$. In
particular, if $tr(\rho)=1$, then $\rho$ is called a density
operator. A (mixed) state of a quantum system is represented by a
density operator. An operator $M$ is said to be Hermitian if its
conjugate operator $M^{\dag}=M$. An observable on a quantum system
is described by a Hermitian operator. A quantum predicate was
defined by D'Hondt and Panangaden~\cite{DP06} to be an observable
$\sqsubseteq I$, where $I$ is the identity operator on $H$.

A main result in this paper depends on the convergence of a certain
sequence of observables, which is guaranteed by the following:

\begin{prop}\label{qdom} (\cite{Se04, K03, YDFJ10}) Both the set of partial density operators
and the set of quantum predicates, equipped with the L\"{o}wner
order, are a CPO (complete partial order).\end{prop}

One of the mathematical formalisms of physically realizable
operations allowed by quantum mechanics is the notion of
super-operator. A super-operator on $H$ is a linear operator
$\mathcal{E}$ from the space of linear operators on $H$ into itself,
which satisfies the following two conditions: \begin{enumerate}
\item $tr[\mathcal{E}(\rho)]\leq tr(\rho)$ for each partial density
operator $\rho$; \item Complete positivity: for any extra Hilbert
space $H_R$, $(\mathcal{I}_R\otimes \mathcal{E})(A)$ is positive
provided $A$ is a positive operator on $H_R\otimes H$, where
$\mathcal{I}_R$ is the identity operation on $H_R$.\end{enumerate}
If condition 1 is strengthened to $tr[\mathcal{E}(\rho)]=tr(\rho)$
for all partial density operators $\rho$, then $\mathcal{E}$ is said
to be trace-preserving. The following theorem gives an elegant
representations of super-operators.

\begin{thm}\label{kraus} (\cite{NC00}, Theorem 8.1) Kraus Operator-Sum Representation:
 $\mathcal{E}$ is a super-operator on $H$ if and only if there exists a set of
operators $\{E_i\}$ satisfying: \begin{enumerate}\item
$\mathcal{E}(\rho)=\sum_{i}E_i\rho E_i^{\dag}$ for all density
operators $\rho$, \item $\sum_{i}E_i^{\dag}E_i\sqsubseteq I$, with
equality for trace-preserving $\mathcal{E}$, where $I$ is the
identity operator on $H$.\end{enumerate}\end{thm}

The L\"{o}wner order induces a partial order between super-operators
in a natural way: $\mathcal{E}\sqsubseteq\mathcal{F}$ if and only if
$\mathcal{E}(\rho)\sqsubseteq\mathcal{F}(\rho)$ for all partial
density operators $\rho$. The convergence of a certain sequence of
super-operators is required in the proof of a key lemma in this
paper. It is guaranteed by the following:

\begin{prop}\label{conve} (\cite{Se04, YDFJ10}) The set of (not necessarily
trace-preserving) super-operators equipped with $\sqsubseteq$ is a
CPO (complete partial order).
\end{prop}

\subsection{Schr\"odinger-Heisenberg Duality}\label{SHD}

Now we can introduce a duality between states described as density operators and observables described as
Hermitian operators. This duality will be our main tool in developing the Sharir-Pnueli-Hart method for
verifying quantum programs.

\begin{defn}Let $\mathcal{E}$ be a super-operator mapping (partial) density operators to (partial) density
operators, and let $\mathcal{E}^{\ast}$ be a super-operator mapping
Hermitian operators to Hermitian operators. If we have
$$tr[M\mathcal{E}(\rho)]=tr[\mathcal{E}^{\ast}(M)\rho]$$ for any
(partial) density operator $\rho$, and for any Hermitian operator
$M$, then we say that $\mathcal{E}$ and $\mathcal{E}^{\ast}$ are
(Schr\"odinger-Heisenberg) dual.
\begin{figure}[h]\centering
$$\begin{array}{ccccc}\rho & \stackrel{\mathcal{E}}{\longrightarrow} & \mathcal{E}(\rho)
\\ \\ \Updownarrow & tr[\mathcal{E}^{\ast}(M)\rho]  =  tr[M\mathcal{E}(\rho)] & \Updownarrow \\ \\ \mathcal{E}^{\ast}(M)
 & \stackrel{\mathcal{E}^{\ast}}{\longleftarrow} & M\end{array}$$

The mapping $\rho\mapsto\mathcal{E}(\rho)$ is the Schr\"{o}dinger
picture, and the mapping $M\mapsto\mathcal{E}^{\ast}(M)$ is the
Heisenberg picture.
 \caption{Schr\"odinger-Heisenberg Duality}\label{fig 1}
\end{figure}
\end{defn}

The following proposition gives an operator-sum representation of the dual of a super-operator, which will be frequently used in the sequent sections.

\begin{prop}If $\mathcal{E}$ has the operator-sum representation $$\mathcal{E}(\rho)=\sum_i E_i\rho E_i^{\dag}$$ for all density operators $\rho$, then we have: $$\mathcal{E}^{\ast}(M)=\sum_i E_i^{\dag} ME_i$$ for all Hermitian operators $M$.\end{prop}

\textit{Proof:} It suffices to see that for any density operator $\rho$, by definition we have:\begin{equation*}\begin{split}
tr[\mathcal{E}^{\ast}(M)\rho]&=tr[M\mathcal{E}(\rho)]=tr[M(\sum_iE_i\rho E_i^{\dag})]\\ &=\sum_itr(ME_i\rho E_i^{\dag})=\sum_i tr(E_i^{\dag}ME_i\rho)\\ &=tr[(\sum_i E_i^{\dag}ME_i)\rho].\ \Box
\end{split}\end{equation*}

The duality considered in this subsection was also exploited by
D'Hondt and Panangaden~\cite{DP06} to establish an elegant
connection between the state transformer (forward) semantics and the
predicate transformer (backward) semantics of quantum programs.

\subsection{Matrix Representation of Super-Operators}

The matrix representation of a super-operator is usually easier to
manipulate than the super-operator itself. This technique will be
extensively used in Sec.~\ref{finite1} and~\ref{finite2}.
\begin{defn}\label{mrd}
Suppose super-operator $\mathcal{E}$ on a finite-dimensional Hilbert
space $H$ has the operator-sum representation:
$$\mathcal{E}(\rho)=\sum_iE_i\rho E_i^{\dag}$$ for all partial density
operators $\rho$, and $\dim H=d$. Then the matrix representation of
$\mathcal{E}$ is the following $d^2\times d^2$ matrix: $$M=\sum_i
E_i\otimes E_i^*,$$ where $A^{\ast}$ stands for the conjugate of
matrix $A$, i.e. $A^{\ast}=(a^{\ast}_{ij})$ with $a^{\ast}_{ij}$
being the conjugate of complex number $a_{ij}$, whenever
$A=(a_{ij})$.
\end{defn}

The next is a key lemma for the proofs of the main results in
Sec.~\ref{finite1} and~\ref{finite2}. Also, it is easy to see from
the next lemma that the matrix representation of a super-operator is
well-defined: if super-operator $\mathcal{E}$ has another
operator-sum representation: $$\mathcal{E}(\rho)=\sum_jF_j\rho
F_j^{\dag}$$ for all partial density operator $\rho$, then
$$\sum_iE_i\otimes E_i^{\ast}=\sum_jF_j\otimes F_j^{\ast}.$$

\begin{lem}\label{mr-lem} We write
$|\Phi\rangle=\sum_j|jj\rangle$ for the (unnormalized) maximally
entangled state in $H\otimes H$, where $\{|j\rangle\}$ is an
orthonormal basis of $H$. Let $M$ be the matrix representation of
super-operator $\mathcal{E}$. Then for any $d\times d$ matrix $A$,
we have: $$(\mathcal{E}(A)\otimes I)|\Phi\rangle=M(A\otimes
I)|\Phi\rangle.$$\end{lem}

\textit{Proof:} We first observe the following matrix equality: for
any matrices $A,B$ and $C$,
\begin{equation*}\label{meq}(A\otimes B)(C\otimes I)|\Phi\rangle=(ACB^T\otimes
I)|\Phi\rangle,\end{equation*} where $B^T$ stands for the transpose
of matrix $B$. Now it follows that
\begin{equation*}\begin{split}
&M(A\otimes I)|\Phi\rangle
=\sum_{i}(E_i\otimes E_i^*)(A\otimes I)|\Phi\rangle\\
&=\sum_{i}(E_iAE_i^{\dag}\otimes
I)|\Phi\rangle=(\mathcal{E}(A)\otimes I)|\Phi\rangle.\
\Box\end{split}
\end{equation*}

\section{Quantum Generalization of Sharir-Pnueli- Hart Method}\label{main}

We first define the model of quantum programs. Let $H$ be a
separable Hilbert space. It will be regarded as the state space of
quantum programs. A quantum program is considered as a quantum
operation on $H$, which transforms a (mixed) state in $H$ to
another. Following Selinger~\cite{Se04}, a single step of the
program may be depicted by a trace-preserving super-operator
$\mathcal{E}$ on $H$. Suppose that $\rho_0$ is the initial state. An
execution of the program starts in $\rho_0$ and repeatedly
implements the quantum operation represented by $\mathcal{E}$.
Formally, we introduce:

\begin{defn}A quantum Markov chain is a triple
$(H,\mathcal{E},\rho_0)$, where:\begin{enumerate}\item $H$ is a
(separable) Hilbert space;\item $\mathcal{E}$ is a trace-preserving
super-operator; and \item $\rho_0$ is a density operator.
\end{enumerate}
\end{defn}

In this paper, since the state space $H$ is fixed, we simply say
that the pair $(\mathcal{E},\rho_0)$ is a quantum Markov chain.

We assume that the program has a terminal (absorbing) space. At the
end of each execution step, we check whether the program reaches a
terminal state or not. This is modeled by a yes-no measurement
$\{M_0,M_1\}$: if the outcome of the measurement is $0$, then the
program terminates, and we can imagine the program state falls into
a terminal space and it remains there forever; otherwise, the
program will enter the next step and continues to perform the
quantum operation $\mathcal{E}$. So, the execution can be seen as a
quantum Markov chain $(\mathcal{E},\rho_0)$ together with a
termination test $\{M_0,M_1\}$. The program is essentially a
generalization of the quantum loop considered in~\cite{YF10} so that
the loop body is allowed to be a super-operator.

To simplify the presentation, we define super-operator:
\begin{equation}\label{super01}\mathcal{E}_i(\rho)=M_i\rho M_i^{\dag}\end{equation} for any
partial density operator $\rho$, and $i=0,1$. Then the execution of
the program can be more precisely described as follows. At the first
step, we perform the termination measurement $\{M_0,M_1\}$ on the
initial state $\rho_0$. The probability that the program terminates;
that is, the measurement outcome is $0$, is
$p_1=tr[\mathcal{E}_0(\rho_0)],$ and the program state after
termination is $\rho_1=\mathcal{E}_0(\rho_0)/p_1.$ We adopt
Selinger's normalization convention~\cite{Se04} to encode
probability $p_1$ and density operator $\rho_1$ into a partial
density operator $p_1\rho_1=\mathcal{E}_0(\rho_0).$ So,
$\mathcal{E}_0(\rho_0)$ is the partial terminal state obtained at
the first step. On the other hand, the probability that the program
does not terminate; that is, the measurement outcome  is $1$, is
$p_1^{N}=tr[\mathcal{E}_1(\rho_0)],$ and the program state after the
outcome $1$ is obtained is
$\rho_1^{N}=\mathcal{E}_1(\rho_0)/p_1^{N}.$ Then they are combined
to get a partial density operator
$p_1^{N}\rho_1^{N}=\mathcal{E}_1(\rho_0),$ and it is transformed by
the defining super-operator of the quantum Markov chain to
$(\mathcal{E}\circ\mathcal{E}_1)(\rho_0)$, upon which the second
step will be executed. In general, the $(n+1)$th step is executed
upon the partial density operator
$$p_{n}^{N}\rho_{n}^{N}=(\mathcal{E}\circ
\mathcal{E}_1)^{n}(\rho),$$ where $p_{n}^{N}$ is the probability
that the program does not terminate at the $n$th step, and
$\rho_{n}^{N}$ is the program state when the program does not
terminate at the $n$th step. The probability that the program
terminates in the $(n+1)$th step is then
\begin{equation}\label{prd1}p_{n+1}=tr([\mathcal{E}_0\circ
(\mathcal{E}\circ\mathcal{E}_1)^{n}](\rho_0)),\end{equation} and the
probability that the program does not terminate in the $(n+1)$th
step is then
\begin{equation}\label{prd2}p^{N}_{n+1}=tr([\mathcal{E}_1\circ
(\mathcal{E}\circ\mathcal{E}_1)^{n}](\rho_0)).\end{equation} The
program state after the termination is
$$\rho_{n+1}=[\mathcal{E}_0\circ
(\mathcal{E}\circ\mathcal{E}_1)^{n}](\rho_0)/p_{n+1},$$ and
combining them yields the partial terminal state of the program at
the $(n+1)$th step: $$p_{n+1}\rho_{n+1}=[\mathcal{E}_0\circ
(\mathcal{E}\circ\mathcal{E}_1)^{n}](\rho_0).$$ Thus, the (total)
terminal state of the program is obtained by summing up the partial
computing results at all steps. Formally, it is given by
$$\rho^{\ast}=\sum_{n=0}^{\infty}[\mathcal{E}_0\circ
(\mathcal{E}\circ \mathcal{E}_1)^{n}](\rho_0).$$ Furthermore, we can
understand that a super-operator $\mathcal{E}$ together with a
termination measurement $\{M_0,M_1\}$ defines a quantum program
scheme. Whenever an initial state $\rho_0$ is fed into the program
scheme, we get a program and it computes the terminal state
$\rho^{\ast}$. Thus, the program scheme transforms initial state
$\rho_0$ to terminal state $\rho^\ast$, and it defines a
super-operator $\mathcal{F}:\rho_0\mapsto\rho^\ast$.

The following proposition gives a recursive characterization of quantum program.

\begin{prop}\label{recur}\begin{enumerate}\item The super-operator $\mathcal{F}$ satisfies the following recursive equation:
\begin{equation}\label{forw}\mathcal{F}(\rho)=\mathcal{E}_0(\rho)+\mathcal{F}[(\mathcal{E}\circ \mathcal{E}_1)(\rho)]
\end{equation} for all density operators $\rho$.
\item The dual $\mathcal{F}^{\ast}$ of $\mathcal{F}$ is defined by the following equation:
\begin{equation}\label{back}\mathcal{F}^{\ast}(M)=\mathcal{E}^{\ast}_0(M)+(\mathcal{E}_1^{\ast}\circ\mathcal{E}^{\ast})[\mathcal{F}^{\ast}(M)]\end{equation} for all Hermitian operators $M$.\end{enumerate}
\end{prop}

\textbf{Verification Problem for Quantum Programs:} Now we consider
a positive observable modeled by Hermitian operator $P\sqsupseteq
0$, where $0$ is the zero operator on $H$. As will be shown in the
examples below, a class of interesting properties of the quantum
program can be expressed as the average value (or expectation)
$$\langle P\rangle_{\rho^{\ast}} \stackrel{def}{=}tr(P\rho^{\ast})$$ of observable $P$
in the terminal state $\rho^{\ast}$ of the program. Our question is then: how to compute it?

The main aim of this paper is to develop a verification method in the Sharir-Pnueli-Hart style to solve this problem.

\textbf{Solving the Problem:} Recall that in the probabilistic case,
a nonnegative vector $\beta=(\beta_j)_{j\in T}$ is given over the
set $T$ of terminal states, and the task is to compute its
expectation with respect to the terminal distribution of the
program. The Sharir-Pnueli-Hart method requires to find a completing
$\overline{\beta}=(\overline{\beta}_i)_{i\in S}$ over the whole set
$S$ of program states satisfying conditions (V1), (V2) and (V3).

A unexpected difficulty comes out when we try to generalize this
method to the quantum case: The notion that a vector
$\overline{\beta}$ is a completion of another vector $\beta$ is
ready and simple. The quantum analogue of vector $\beta$ is the
positive operator $P$. What is the quantum analogue of the
completion $\overline{\beta}$ of $\beta$? Note that $\beta$ is
defined merely over a subset $T$ of $S$, and $\overline{\beta}$ can
be obtained by adding its values in the remaining states in
$S\setminus T$. However, $P$ is defined already on the whole space
$H$ of program states but not one of its subspaces. So, at the first
glance it seems hopeless to define a completion of $P$ which must
also be an operator on $H$. After a careful analysis, however, we
understand that $$M_0^{\dag}PM_0+M_1^{\dag}QM_1$$ with a positive
operator $Q$ on $H$ can be seen as a completion of $P$ with respect
to the termination measurement $\{M_0,M_1\}$. Once a suitable
completion of $P$ is discovered, it is relatively easy to conceive
the quantum counterparts of conditions (V1), (V2) and (V3). They are
displayed as follows:
\begin{itemize}
\item (QV1) $\langle M_0^{\dag}PM_0+M_1^{\dag}QM_1\rangle_{\rho_0}<\infty$;
\item (QV2) $\mathcal{E}^{\ast}(M_0^{\dag}PM_0+M_1^{\dag}QM_1)=Q$;
\item (QV3) ($Q-$termination) $$\lim_{n\rightarrow\infty}tr(Q[\mathcal{E}_1\circ
(\mathcal{E}\circ \mathcal{E}_1)^{n}](\rho_0))=0,$$\end{itemize}

It is interesting to compare (QV2) with (V2). Note that in (QV2) the
dual $\mathcal{E}^\ast$ but not $\mathcal{E}$ itself is used. This
shows that the Schr\"{o}dinger-Heisenberg duality must be taken
seriously for quantum programs. But it is not the case in condition
(V2) for probabilistic programs.

Recall that the Sharir-Pnueli-Hart method is based on the
implication from (V1), (V2) and (V3) to (C). Similarly, if (QV1),
(QV2) and (QV3) imply
\begin{itemize}
\item (QC) $\langle P\rangle_{\rho^{\ast}}=\langle
M_0^{\dag}PM_0+M_1^{\dag}QM_1\rangle_{\rho_0}$,
\end{itemize}then its quantum generalization can be
stated as follows:

\begin{itemize}\item
\textit{\textbf{Quantum Sharir-Pnueli-Hart method}: To compute the
average value $\langle P\rangle_{\rho^{\ast}}$ of positive
observable $P$ in the terminal state $\rho^{\ast}$ of the program,
one only needs to find a positive operator $Q$ satisfying conditions
(QV1), (QV2) and (QV3). Then the problem of computing $\langle
P\rangle_{\rho^{\ast}}$ is reduced to computing the average value
$\langle M_0^{\dag}PM_0+M_1^{\dag}QM_1\rangle_{\rho_0}$ of
observable $M_0^{\dag}PM_0+M_1^{\dag}QM_1$ in the initial state
$\rho_0$ of the program.}\end{itemize}

It is worth noting that $P$ is assumed to be positive in the above
presentation of quantum Sharir-Pnueli-Hart method. The positivity of
$P$ is essential in the proofs of the correctness and completeness
theorems below, where the monotonicity of certain sequences of
(super-)operators is crucial. In practical applications, however,
the positivity of $P$ is not essential. In fact, for any observable
(Hermitian operator) $O$, it holds that $O=P_1-P_2$ for some
positive operators $P_1$ and $P_2$. Thus, we can use the above
quantum Sharir-Pnueli-Hart method to compute $\langle
P_1\rangle_{\rho^\ast}$ and $\langle P_2\rangle_{\rho^\ast}$,
respectively, and we obtain: $$\langle O\rangle_{\rho^\ast}=\langle
P_1\rangle_{\rho^\ast}-\langle P_2\rangle_{\rho^\ast}$$ provided
$\langle P_1\rangle_{\rho^\ast}\neq\infty$ or $\langle
P_2\rangle_{\rho^\ast}\neq\infty$.

We postpone the proof of the correctness of the quantum
Sharir-Pnueli-Hart method to a separate subsection, but now present
a simple example to show how to use it.

\begin{exam}\label{ex1}
We consider the bit flip channel, which is widely used in quantum
communication. This channel flips the state of a qubit from
$|0\rangle$ to $|1\rangle$ and vice versa, with probability $1-p$,
$0\leq p\leq 1$ (see~\cite{NC00}, Sec. 8.3.3). It is described by
the super-operator $\mathcal{E}$ on the $2-$dimensional Hilbert
space $H_2$, defined as follows: $$\mathcal{E}(\rho)=E_0\rho
E_0+E_1\rho E_1$$ for all density operator $\rho$, where
$E_0=\sqrt{p}I,$ and $E_1=\sqrt{1-p}X.$ Assume that the initial
state is $\rho_{0}=|\psi_{0}\rangle\langle \psi_{0}|$ with
$|\psi_{0}\rangle=\alpha |0\rangle+\beta |1\rangle,$ and the
termination measurement $M=\{M_0,M_1\}$ is the measurement in the
computational basis, i.e. $M_0=|0\rangle\langle 0|,\ \ \
M_1=|1\rangle\langle 1|.$

Let us compute the termination probability of the program defined by
super-operator $\mathcal{E}$, initial state $\rho_0$ and termination
measurement $M$. Then we take $P=|0\rangle\langle 0|,$ and the
termination probability is $\langle P\rangle_{\rho^{\ast}}$, where
$\rho^{\ast}$ is the terminal state of the program. To use the
quantum Sharir-Pnueli-Hart method, we need to find a positive
operator $Q$ satisfying condition (QV2). It is reasonable to assume
that $Q=\lambda |\varphi\rangle\langle\varphi|+\mu
|\psi\rangle\langle\psi|$ such that
$\{|\varphi\rangle,|\psi\rangle\}$ is an orthonormal basis of $H_2$,
and $\lambda,\mu\geq 0$. We write:
$$N=M_0^{\dag}PM_0+M_1^{\dag}QM_1.$$ A routine calculation leads to
$N=|0\rangle\langle 0|+K|1\rangle\langle 1|,$ and
$$\mathcal{E}^{\ast}(N)=[p+(1-p)K]|0\rangle\langle 0|+[pK+(1-p)]|1\rangle\langle 1|$$
where and in the sequel $$K=\lambda |\langle\varphi |1\rangle
|^{2}+\mu |\langle\psi|1\rangle|^{2}.$$ Then condition (QV2) becomes
$\mathcal{E}^{\ast}(N)=Q$, and it yields that
\begin{equation}\label{kk}K=\langle 1|Q|1\rangle=\langle 1|\mathcal{E}^{\ast}(N)|1\rangle=pK+(1-p).\end{equation}

On the other hand, by a routine calculation we obtain:
$$[\mathcal{E}_1\circ (\mathcal{E}\circ
\mathcal{E}_1)^{n}](\rho^{0})=|\beta|^{2}p^{n}|1\rangle\langle1|,$$
$$tr(Q[\mathcal{E}_1\circ (\mathcal{E}\circ
\mathcal{E}_1)^{n}](\rho^{0}))=|\beta|^{2}Kp^{n}.$$

We consider the following three cases:
\begin{itemize}\item
Case 1. $p<1$. We always have (QV3). It follows from Eq.~(\ref{kk})
that $K=1$. So, $N=I$, and the assertion (QC) implies the
termination probability $\langle P\rangle_{\rho^{\ast}}=\langle
N\rangle_{\rho_{0}}=\langle\psi_{0}|N|\psi_{0}\rangle=1.$
\item
Case 2. $\beta=0$. Again, condition (QV3) automatically holds. Then
$|\psi_{0}\rangle=e^{i\theta}|0\rangle$ for some real number
$\theta$, and by (QC) we obtain the termination probability $\langle
P\rangle_{\rho^{\ast}}=\langle\psi_{0}|N|\psi_{0}\rangle=1.$
\item
Case 3. $\beta\neq 0$ and $p=1$. To guarantee (QV3), we must take
$K=0$. Then $N=|0\rangle\langle 0|,$ and the  termination
probability $\langle
P\rangle_{\rho^{\ast}}=\langle\psi_{0}|N|\psi_{0}\rangle=|\alpha|^{2}.$\end{itemize}
\end{exam}

\subsection{Correctness Theorem}

This subsection is devoted to prove the correctness of the quantum
Sharir-Pnueli-Hart method:

\begin{thm}\label{corr}(Correctness) The quantum Sharir-Pnueli-Hart method is correct; that is, (QV1), (QV2) and (QV3) imply (QC).\end{thm}

A key intermediate step in the proof of the above theorem is the
calculation of the expectation in the right-hand side of (QC). It
can be stated as the following:

\begin{lem}\label{k-lem}If $Q$ satisfies (QV2), then for any $n\geq 0$, we have:
\begin{equation}\label{ind}\begin{split}
\langle M_0^{\dag}PM_0+&M_1^{\dag}QM_1\rangle_{\rho_0}
=\sum_{k=0}^{n}tr(P[\mathcal{E}_0\circ (\mathcal{E}\circ
\mathcal{E}_1)^{k}](\rho_0))\\ &\ \ \ \ \ \ +tr(Q[\mathcal{E}_1\circ
(\mathcal{E}\circ \mathcal{E}_1)^{n}](\rho_0)).\end{split}
\end{equation}
\end{lem}

\textit{Proof.} The key of this proof is repeated applications of
the Schr\"{o}dinger-Heisenberg duality between states and
observables defined in Subsec.~\ref{SHD}. We proceed by induction on
$n$. For the case of $n=0$, it holds that
\begin{equation*}\begin{split}
RHS&=tr(PM_0\rho_0M_0^{\dag})+tr(QM_1\rho_0M_1^{\dag})\\
&=tr[(M_0^{\dag}PM_0+M_1^{\dag}QM_1)\rho_0]=LHS.
\end{split}\end{equation*}We now consider the case of $n+1$. First, by (QV2) and the duality between $\mathcal{E}$ and $\mathcal{E}^{\ast}$, we have:
\begin{equation*}\begin{split}
&\mathcal{R}\stackrel{def}{=}tr(P[\mathcal{E}_0\circ (\mathcal{E}\circ \mathcal{E}_1)^{n+1}(\rho_0))\\ &\ \ \ \ \ \ \ \ \ \ \ \ \ \ \ \ \ \ \ \ \ \ \ \ \ \ \ \ \ \ \ \ \ \ \ +tr(Q[\mathcal{E}_1\circ (\mathcal{E}\circ \mathcal{E}_1)^{n+1}](\rho_0))\\
&=tr[PM_0(\mathcal{E}\circ \mathcal{E}_1)^{n+1}(\rho_0)M_0^{\dag}]\\ &\ \ \ \ \ \ \ \ \ \ \ \ \ \ \ \ \ \ \ \ \ \ \ \ \ \ \ \ \ \ \ \ \ \ \ +tr[QM_1 (\mathcal{E}\circ \mathcal{E}_1)^{n+1}(\rho_0)M_1^{\dag}]\\
&=tr[(M_0^{\dag}PM_0+M_1^{\dag}QM_1)(\mathcal{E}\circ\mathcal{E}_1)^{n+1}(\rho_0)]\\
&=tr\{(M_0^{\dag}PM_0+M_1^{\dag}QM_1)\mathcal{E}([\mathcal{E}_1\circ (\mathcal{E}\circ\mathcal{E}_1)^{n}](\rho_0))\}\\
&=tr\{\mathcal{E}^{\ast}(M_0^{\dag}PM_0+M_1^{\dag}QM_1)[\mathcal{E}_1\circ (\mathcal{E}\circ\mathcal{E}_1)^{n}](\rho_0)\}\\
&=tr\{Q[\mathcal{E}_1\circ
(\mathcal{E}\circ\mathcal{E}_1)^{n}](\rho_0)\}.
\end{split}\end{equation*} Then it follows from the induction hypothesis for the case of $n$ that \begin{equation*}\begin{split}
RHS&=\sum_{k=0}^{n+1}tr(P[\mathcal{E}_0\circ (\mathcal{E}\circ \mathcal{E}_1)^{k}](\rho_0))\\ &\ \ \ \ \ \ \ \ \ \ \ \ \ \ \ \ \ \ \ \ \ \ \ \ \ \ \ \ \ \ +tr(Q[\mathcal{E}_1\circ (\mathcal{E}\circ \mathcal{E}_1)^{n+1}](\rho_0))\\
&=\sum_{k=0}^{n}tr(P[\mathcal{E}_0\circ (\mathcal{E}\circ \mathcal{E}_1)^{k}](\rho_0))+\mathcal{R}\\
&=\sum_{k=0}^{n}tr(P[\mathcal{E}_0\circ (\mathcal{E}\circ
\mathcal{E}_1)^{k}](\rho_0))\\ &\ \ \ \ \ \ \ \ \ \ \ \ \ \ \ \ \ \
\ \ \ \ \ \ \ \ \ \ \ \ +tr(Q[\mathcal{E}_1\circ (\mathcal{E}\circ
\mathcal{E}_1)^{n}](\rho_0))\\ &=LHS.\ \
\Box\end{split}\end{equation*}

The above lemma will also be needed in the proof of the completeness
theorem below. Now we are ready to prove the correctness theorem.

\textit{Proof of Theorem~\ref{corr}.} Combining Lemma~\ref{k-lem}
and conditions (QV1), (QV2) and (QV3), we obtain:

\begin{equation*}\begin{split}\langle
M_0^{\dag}&PM_0+M_1^{\dag}QM_1\rangle_{\rho_0}\\
&=\lim_{n\rightarrow\infty}\sum_{k=0}^{n}tr(P[\mathcal{E}_0\circ
(\mathcal{E}\circ\mathcal{E}_1)^{k}](\rho_0))\\
&=\lim_{n\rightarrow\infty}tr(\sum_{k=0}^{n}P[\mathcal{E}_0\circ
(\mathcal{E}\circ\mathcal{E}_1)^{k}](\rho_0))\\
&=\lim_{n\rightarrow\infty}tr(P\sum_{k=0}^{n}[\mathcal{E}_0\circ
(\mathcal{E}\circ\mathcal{E}_1)^{k}](\rho_0))\\
&\stackrel{(a)}{=}tr(\lim_{n\rightarrow\infty}P\sum_{k=0}^{n}[\mathcal{E}_0\circ
(\mathcal{E}\circ\mathcal{E}_1)^{k}](\rho_0))\\
&\stackrel{(b)}{=}tr(P\lim_{n\rightarrow\infty}\sum_{k=0}^{n}[\mathcal{E}_0\circ
(\mathcal{E}\circ\mathcal{E}_1)^{k}](\rho_0))\\
&=tr(P\rho^{\ast})=\langle P\rangle_{\rho^{\ast}}.
\end{split}\end{equation*}Note that the existence of the limits in the above equation
is guaranteed by the fact that the set of partial density operators
equipped with the L\"{o}wner order is a CPO (see
Proposition~\ref{qdom}), and the equalities labeled by (a) and (b)
are derived by the continuity of trace and the continuity of
multiplication of operators, respectively. $\Box$

\subsection{Completeness Theorem}

The quantum Sharir-Pnueli-Hart method was used in the last section
to compute the termination probability of a simple quantum program
used in quantum communication. It is natural to ask the question:
Can it be used to more complicated quantum programs? Furthermore,
one may ask: Is the quantum Sharir-Pnueli-Hart method complete? More
precisely, for any positive observable $P$, can we always find a
positive observable $Q$ such that we only need to compute the
average value of observable $M_0^{\dag}PM_0+M_1^{\dag}QM_1$ in the
initial state $\rho_0$ in order to compute the average value of $P$
in the terminal state $\rho^{\ast}$? The following theorem gives a
positive answer to this question:

\begin{thm} (Completeness) For any positive observable $P$, there is
always a positive observable $Q$ satisfying condition (QV2), and
\begin{enumerate}
\item if $Q$ satisfies (QV1), then it also satisfies (QV3) and thus
(QC);\item if it does not satisfies (QV1), then $\langle
P\rangle_{\rho^{\ast}}=\infty$.
\end{enumerate}Indeed, $Q$ can be chosen to be the least fixed-point of the following
equation:
\begin{equation}\label{fix}\mathcal{E}^{\ast}(M_0^{\dag}PM_0+M_1^{\dag}QM_1)=Q,\
Q\geq 0.\end{equation}
\end{thm}

\textit{Proof.} We put: $Q_0=0,$ and
$$Q_{n+1}=M_0^{\dag}PM_0+M_1^{\dag}\mathcal{E}^{\ast}(Q_n)M_1,\
n\geq 0.$$ It can be easily shown by induction on $n$ that $\{Q_n\}$
is an increasing sequence according to the L\"{o}wner order. Thus,
we can define:
$$\overline{Q}=\mathcal{E}^{\ast}(\bigvee_{n=0}^{\infty}Q_n).$$

(1) We first show that $\overline{Q}$ is well-defined. It suffices
to prove the existence of supremum $\bigvee_{n=0}^{\infty}Q_n$. With
its monotonicity, we only need to demonstrate that sequence
$\{Q_n\}$ is bounded from up. Indeed, we have:
$Q_n\sqsubseteq\mathcal{F}^{\ast}(P)$ for every $n\geq 0.$ We prove
this claim by induction on $n$. The case of $n=0$ is obvious.
Assuming the claim is correct for the case $n$, we obtain: for any
density operator $\rho$,
\begin{equation*}\begin{split}
&tr(Q_{n+1}\rho)=tr[(M_0^\dag
PM_0+M_1^\dag\mathcal{E}^\ast(Q_n)M_1)\rho]\\&\stackrel{(a)}{=}tr(PM_0\rho
M_0^\dag)+tr[Q_n\mathcal{E}(M_1\rho M_1^\dag)]\\
&=tr(P\mathcal{E}_0(\rho))+tr[Q_n(\mathcal{E}\circ\mathcal{E}_1)(\rho)]\\
&\stackrel{(b)}{\leq}
tr(P\mathcal{E}_0(\rho))+tr[\mathcal{F}^\ast(P)(\mathcal{E}\circ\mathcal{E}_1)(\rho)]\\
&\stackrel{(c)}{=}tr(P\mathcal{E}_0(\rho)+P\mathcal{F}[(\mathcal{E}\circ\mathcal{E}_1)(\rho)])\\
&=tr(P(\mathcal{E}_0(\rho)+\mathcal{F}[(\mathcal{E}\circ\mathcal{E}_1)(\rho)]))\\
&\stackrel{(d)}{=}tr(P\mathcal{F}(\rho))=tr(\mathcal{F}^\ast(P)\rho).
\end{split}\end{equation*}Here, equalities labeled by (a) and (c)
are derived by the Schr\"{o}dinger-Heisenberg duality, (b) by the
induction hypothesis, and (d) by Eq.~(\ref{forw}). So,
$Q_{n+1}\sqsubseteq\mathcal{F}^\ast(P)$, and the claim is proved.

(2) Second, we prove that $\overline{Q}$ satisfies (QV2). It
suffices to show that
$$tr[\mathcal{E}^{\ast}(M_0^{\dag}PM_0+M_1^{\dag}\overline{Q}M_1)\rho]=tr(\overline{Q}\rho)$$ for any density operator $\rho$. The
key is also repeated applications of the Schr\"{o}dinger-Heisenberg
duality. It follows from continuity of trace operator $tr(\cdot)$
that
\begin{equation*}\begin{split}
&LHS=tr[(M_0^{\dag}PM_0+M_1^{\dag}\overline{Q}M_1)\mathcal{E}(\rho)]\\
&=tr[M_0^{\dag}PM_0\mathcal{E}(\rho)]+tr[M_1^{\dag}\overline{Q}M_1\mathcal{E}(\rho)]\\
&=tr[M_0^{\dag}PM_0\mathcal{E}(\rho)]+tr[\overline{Q}M_1\mathcal{E}(\rho)M_1^{\dag}]\\
&=tr[M_0^{\dag}PM_0\mathcal{E}(\rho)]+tr[(\bigvee_{n=0}^{\infty}Q_n)\mathcal{E}(M_1\mathcal{E}(\rho)M_1^{\dag})]\\
&=tr[M_0^{\dag}PM_0\mathcal{E}(\rho)]+\bigvee_{n=0}^{\infty}tr[Q_n\mathcal{E}(M_1\mathcal{E}(\rho)M_1^{\dag})]\\
&=tr[M_0^{\dag}PM_0\mathcal{E}(\rho)]+\bigvee_{n=0}^{\infty}tr[\mathcal{E}^{\ast}(Q_n)M_1\mathcal{E}(\rho)M_1^{\dag}]\\
&=tr[M_0^{\dag}PM_0\mathcal{E}(\rho)]+\bigvee_{n=0}^{\infty}tr[M_1^{\dag}\mathcal{E}^{\ast}(Q_n)M_1\mathcal{E}(\rho)]\\
&=\bigvee_{n=0}^{\infty}tr[(M_0^{\dag}PM_0+M_1^{\dag}\mathcal{E}^{\ast}(Q_n)M_1)\mathcal{E}(\rho)]\\
&=\bigvee_{n=0}^{\infty}tr[Q_{n+1}\mathcal{E}(\rho)]
=tr[(\bigvee_{n=0}^{\infty}Q_{n+1})\mathcal{E}(\rho)]\\
&=tr[(\bigvee_{n=0}^{\infty}Q_{n})\mathcal{E}(\rho)] =RHS.
\end{split}\end{equation*}

(3) Third, we show that $\overline{Q}$ is the least fixed-point of
Eq.~(\ref{fix}). Suppose that $Q$ satisfies (QV2). First, we show
that $tr[Q_n\mathcal{E}(\rho)]\leq tr(Q\rho)$ for any density
operator $\rho$ and for all $n\geq 0$ by induction on $n$. The case
of $n=0$ is trivial. If the conclusion is correct for the case of
$n$, then we have:\begin{equation*}\begin{split}
&tr[Q_{n+1}\mathcal{E}(\rho)]=tr[(M_0^{\dag}PM_0+M_1^{\dag}\mathcal{E}^{\ast}(Q_n)M_1)\mathcal{E}(\rho)]\\
&=tr[M_0^{\dag}PM_0\mathcal{E}(\rho)]+tr[M_1^{\dag}\mathcal{E}^{\ast}(Q_n)M_1\mathcal{E}(\rho)]\\
&=tr[M_0^{\dag}PM_0\mathcal{E}(\rho)]+tr[Q_n\mathcal{E}(M_1\mathcal{E}(\rho)M_1^{\dag})]\\
&\leq tr[M_0^{\dag}PM_0\mathcal{E}(\rho)]+tr[QM_1\mathcal{E}(\rho)M_1^{\dag}]\\
&=tr[(M_0^{\dag}PM_0+M_1^{\dag}QM_1)\mathcal{E}(\rho)]\\
&=tr[\mathcal{E}^{\ast}(M_0^{\dag}PM_0+M_1^{\dag}QM_1)\rho]=tr(Q\rho),
\end{split}\end{equation*}and the conclusion also holds for the case of $n+1$.

Now, we obtain:\begin{equation*}\begin{split}
tr(\overline{Q}\rho)=tr[(\bigvee_{n=0}^{\infty}Q_n)\mathcal{E}(\rho)]
=\bigvee_{n=0}^{\infty}tr[Q_n\mathcal{E}(\rho)]\leq tr(Q\rho)
\end{split}\end{equation*}for all density operator $\rho$. Consequently, $\overline{Q}\leq Q$, and $\overline{Q}$ is the least fixed-point of Eq.~(\ref{fix}).

(4) We claim that \begin{equation}\label{mid}
tr[P\mathcal{E}_0(\rho)+Q_n(\mathcal{E}\circ\mathcal{E}_1)(\rho)]=\sum_{k=0}^{n}tr(P[\mathcal{E}_0\circ
(\mathcal{E}\circ \mathcal{E}_1)^{k}](\rho))
\end{equation}for all density operators $\rho$ and for all $n\geq 0$. To prove Eq.~(\ref{mid}), we proceed by induction on $n$. The case of $n=0$ is obvious. If Eq.~(\ref{mid}) is valid for the case of $n$, then we have:\begin{equation*}\begin{split}
&tr[P\mathcal{E}_0(\rho)+Q_{n+1}(\mathcal{E}\circ\mathcal{E}_1)(\rho)]=tr(P\mathcal{E}_0(\rho))+\\ &\ \ \ tr[P(\mathcal{E}_0\circ(\mathcal{E}\circ \mathcal{E}_1))(\rho)]+tr[\mathcal{E}^{\ast}(Q_n)(\mathcal{E}_1\circ(\mathcal{E}\circ\mathcal{E}_1))(\rho)]\\ &=tr(P\mathcal{E}_0(\rho))+tr[P\mathcal{E}((\mathcal{E}\circ\mathcal{E}_1)(\rho))+Q_n(\mathcal{E}\circ\mathcal{E}_1)^{2}(\rho)]\\
&=tr(P\mathcal{E}_0(\rho))+\sum_{k=0}^{n}tr(P[\mathcal{E}_0\circ(\mathcal{E}\circ\mathcal{E}_1)^{k}]((\mathcal{E}\circ\mathcal{E}_1)(\rho)))\\
&=\sum_{k=0}^{n+1}tr(P[\mathcal{E}_0\circ (\mathcal{E}\circ
\mathcal{E}_1)^{k}](\rho)).
\end{split}\end{equation*}

Now it follows from Eq.~(\ref{mid}) that \begin{equation}\label{mid1}\begin{split}&\langle M_0^{\dag}PM_0+M_1^{\dag}\overline{Q}M_1\rangle_{\rho^{0}}=tr(P\mathcal{E}_0(\rho^{0}))+tr(\overline{Q}\mathcal{E}_1(\rho^{0}))\\ &=tr(P\mathcal{E}_0(\rho^{0}))+tr[(\bigvee_{n=0}^{\infty}Q_n)(\mathcal{E}\circ\mathcal{E}_1)(\rho^{0})]\\ &=\bigvee_{n=0}^{\infty}tr[P\mathcal{E}_0(\rho^{0})+Q_n(\mathcal{E}\circ\mathcal{E}_1)(\rho^{0})]\\ &=\sum_{n=0}^{\infty}tr(P[\mathcal{E}_0\circ(\mathcal{E}\circ\mathcal{E}_1)^{n}](\rho^{0}))\\
&=tr(P\rho^{\ast})=\langle P\rangle_{\rho^{\ast}}.\end{split}
\end{equation}

If $\overline{Q}$ satisfies (QV1), i.e. $\langle
M_0^{\dag}PM_0+M_1^{\dag}\overline{Q}M_1\rangle_{\rho^{0}}<\infty,$
then we derive that $\overline{Q}$ satisfies (QV3), i.e.
$$\lim_{n\rightarrow\infty}(\overline{Q}[\mathcal{E}\circ(\mathcal{E}\circ\mathcal{E}_1)^{n}](\rho^{0}))=0$$
by combining Eq.~(\ref{mid1}) and Lemma~\ref{k-lem}. For the case
that $\overline{Q}$ violates (QV1), Eq.~(\ref{mid1}) implies that
$\langle P\rangle_{\rho^{\ast}}=\infty$.\ \ $\Box$

\section{Verification in Finite-Dimensional State Spaces}\label{finite1}

The Sharir-Pnueli-Hart method developed in the last section can be
used to verify properties of quantum programs both in
finite-dimensional state spaces and in infinite-dimensional state
spaces. In this section, we present a much simpler verification
method for the case of finite-dimensional state spaces.

We consider the program defined by quantum Markov chain
$(\mathcal{E},\rho_0)$ with termination test $\{M_0,M_1\}$ in a
finite-dimensional state Hilbert space $H$. Suppose that
super-operator $\mathcal{E}$ has the operator-sum representation:
$\mathcal{E}=\sum_iE_i\rho E_i^{\dag}$ for all partial density
operators $\rho$. Recall that super-operators $\mathcal{E}_i$
$(i=0,1)$ are defined by Eq.~(\ref{super01}). For simplicity of
presentation, we write $\mathcal{G}$ for the composition of
$\mathcal{E}$ and $\mathcal{E}_1$:
$\mathcal{G}=\mathcal{E}\circ\mathcal{E}_1.$ Then $\mathcal{G}$ has
the following operator-sum representation:
$$\mathcal{G}(\rho)=\sum_iE_iM_1\rho M^{\dag}_1E_i$$ for all partial density operators
$\rho$. Let the matrix representations of super-operators
$\mathcal{E}_0$ and $\mathcal{G}$ are
\begin{equation}\label{mmatrix}\begin{split}
N_i&=M_i\otimes M_i^{\ast},\ i=0,1,\\ M&=\sum_i(E_iM_1)\otimes
(E_iM_1)^{\ast},
\end{split}\end{equation}respectively (see Definition~\ref{mrd}).
Suppose that the Jordan decomposition of $M$ is $M=SJ(M)S^{-1},$
where $S$ is a nonsingular matrix, and $J(M)$ is the Jordan normal
form of $M$: $$J(M)= diag (J_{k_1}(\lambda_1),
J_{k_2}(\lambda_2),\cdot\cdot\cdot, J_{k_l}(\lambda_l))$$ with
$J_{k_s}(\lambda_s)$ being a $k_s\times k_s$-Jordan block of
eigenvalue $\lambda_s$ $(1\leq s\leq l)$ (see~\cite{HJ85}, Sec.
3.1). The next lemma describes the structure of the matrix
representation $M$ of super-operator $\mathcal{F}$.

\begin{lem}\label{tech0}\begin{enumerate}\item $|\lambda_s|\leq 1$ for all $1\leq s\leq l$. \item If $|\lambda_s|=1$
then the dimension of the $s$th Jordan block $k_s=1$.\end{enumerate}
\end{lem}

The verification method of this section heavily depends on the
convergence of power series $\sum_m M^n$ of matrix $M$. But this
series may not converge when some of its eigenvalues has module $1$.
So, we need to modify the Jordan normal form $J(M)$ of $M$ by
vanishing the Jordan blocks corresponding to those eigenvalues with
module $1$, which are all $1-$dimensional according to
Lemma~\ref{tech0}. This yields the matrix: $N=SJ(N)S^{-1},$ where
$$J(N)= diag (J^{\prime}_1,
J^{\prime}_2,\cdot\cdot\cdot, J^{\prime}_3),$$
$$J^{\prime}_s=\begin{cases} 0 &{\rm if}\ |\lambda_s|=1,\\ J_{k_s}(\lambda_s)
&{\rm otherwise,}\end{cases}$$ for each $1\leq s\leq l$. (Note that
$J_{k_s}(\lambda_s)$ and $J^{\prime}_s$ are both $1\times 1$
matrices when $\lambda_s|=1$; see Lemma~\ref{tech0}.)

The following technical lemma is crucial for the proofs of the main
results in this and next section.

\begin{lem}\label{tech}For any integer $n\geq 0$, we have: $N_0M^{n}=N_0N^{n}.$\end{lem}

The lengthy proofs of Lemmas~\ref{tech0} and \ref{tech} are
postponed to the appendix. Now we are ready to present the main
result of this section.

\begin{thm}\label{find}For any Hermitian operator $P$ on the state space $H$,
the expectation of observable $P$ in the terminal state
$\rho^{\ast}$ of the program is $$\langle
P\rangle_{\rho^{\ast}}=\langle\Phi|(P\otimes
I)N_0(I-N)^{-1}(\rho_0\otimes I)|\Phi\rangle,$$ where $I$ is the
identity operator on $H$, and $|\Phi\rangle=\sum_j|jj\rangle$ is the
(unnormalized) maximally entangled state in $H\otimes H$, with
$\{|j\rangle\}$ being an orthonormal basis of $H$.
\end{thm}

\textit{Proof:} It follows from Lemma~\ref{mr-lem} together with the
defining equation of $\mathcal{E}_0$ and $\mathcal{F}$ that
\begin{equation}\label{meq1}[\mathcal{E}_0(\rho)\otimes
I]|\Phi\rangle=N_0(\rho\otimes I)|\Phi\rangle,\end{equation}
\begin{equation}\label{meq2}[\mathcal{F}(\rho)\otimes
I]|\Phi\rangle=M(\rho\otimes I)|\Phi\rangle.\end{equation} By first
applying Eq.~(\ref{meq1}) and then repeatedly applying
Eq.~(\ref{meq2}), we obtain:\begin{equation*}\begin{split}
&(\rho^{\ast}\otimes I)|\Phi\rangle
=[\sum_{n=0}^{\infty}\mathcal{E}_0(\mathcal{F}^{n}(\rho_0))\otimes
I]|\Phi\rangle\\
&=\sum_{n=0}^{\infty}[\mathcal{E}_0(\mathcal{F}^{n}(\rho_0))\otimes
I]|\Phi\rangle\\
&=\sum_{n=0}^{\infty}[\mathcal{E}_0(\mathcal{F}^{n}(\rho_0))\otimes
I]|\Phi\rangle\\
&=\sum_{n=0}^{\infty}N_0(\mathcal{F}^{n}(\rho_0)\otimes
I)|\Phi\rangle =\sum_{n=0}^{\infty}N_0M^{n}(\rho_0\otimes
I)|\Phi\rangle\\
&\stackrel{(a)}{=}\sum_{n=0}^{\infty}N_0N^{n}(\rho_0\otimes
I)|\Phi\rangle
=N_0(\sum_{n=0}^{\infty}N^{n})(\rho_0\otimes I)|\Phi\rangle\\
&=N_0(I-N)^{-1}(\rho_0\otimes I)|\Phi\rangle.
\end{split}\end{equation*}The equality labeled by (a) follows from Lemma~\ref{tech}. Finally, we have:
\begin{equation*}\begin{split}\langle
P\rangle_{\rho^{\ast}}&=tr(P\rho^{\ast})
=\langle\Phi|P\rho^{\ast}\otimes I|\Phi\rangle\\ &=\langle\Phi|(P\otimes I)(\rho^{\ast}\otimes I)|\Phi\rangle\\
&=\langle\Phi|(P\otimes I)N_0(I-N)^{-1}(\rho_0\otimes
I)|\Phi\rangle.\ \Box
\end{split}\end{equation*}

\begin{exam}\label{ex2}We reconsider Example~\ref{ex1}. According to Eq.~(\ref{mmatrix}), put
$N_0=M_0\otimes M_0$ and
$$M=E_0M_1\otimes E_0M_1+E_1M_1\otimes E_1M_1=\left(\begin{array}{cccc}0 & 0 & 0 & 1-p\\ 0 & 0 & 0 & 0\\ 0& 0& 0& 0\\ 0& 0& 0& p\end{array}\right)$$
The case of $p=1$ is trivial. For the case of $p<1$, we have:
$$(I-M)^{-1}=\left(\begin{array}{cccc}1 & 0 & 0 & 1\\ 0 & 1 & 0 & 0\\ 0& 0&
1& 0\\ 0& 0& 0& \frac{1}{1-p}\end{array}\right)$$ To compute the
termination probability of the program, we consider the observable
$P=|0\rangle\langle 0|$ and write
$|\Phi\rangle=|00\rangle+|11\rangle$. Then by Theorem~\ref{find} and
some routine matrix multiplications we obtain the termination
probability:
\begin{equation*}\begin{split}
\langle P\rangle_{\rho^{\ast}}&=\langle\Phi|(P\otimes
I)N_0(I-N)^{-1}(\rho_0\otimes I)|\Phi\rangle\\ &=\langle
0|\rho_0|0\rangle+\langle 1|\rho_0|1\rangle\\ &=tr(\rho_0)=1.
\end{split}\end{equation*}
\end{exam}

To further illustrate the power of the verification method
introduced above, we compute the average running time:
$\sum_{n=1}^{\infty}np_n$ of the program, where $p_n$ is the
probability that the program terminates in the $n$th step (see
Eq.~(\ref{prd1}) for the definition of $p_n$). It is clear that this
cannot be done by a direct application of Theorem~\ref{find}. But a
procedure similar to the proof of Theorem~\ref{find} leads to:

\begin{prop}\label{prop}With the same notation as in Theorem~\ref{find}, the average running time of the program is
$$\langle\Phi|N_0(I-N)^{-2}(\rho_0\otimes I)|\Phi\rangle.
$$\end{prop}

\textit{Proof:} We have: \begin{equation*}\begin{split}
\sum_{n=1}^{\infty}np_n&=\sum_{n=1}^{\infty}n
tr[(\mathcal{E}_0\circ\mathcal{F}^{n-1})(\rho_0)]\\
&=\sum_{n=1}^{\infty}n\langle\Phi|(\mathcal{E}_0\circ\mathcal{F}^{n-1})(\rho_0)\otimes
I|\Phi\rangle\\ &=\sum_{n=1}^{\infty}n\langle\Phi
|N_0M^{n-1}(\rho_0\otimes I)|\Phi\rangle\\
&=\sum_{n=1}^{\infty}n\langle\Phi
|N_0N^{n-1}(\rho_0\otimes I)|\Phi\rangle\\
&=\langle\Phi|N_0(\sum_{n=1}^{\infty}nN^{n-1})(\rho_0\otimes I)|\Phi\rangle\\
&=\langle\Phi|N_0(I-N)^{-2}(\rho_0\otimes I)|\Phi\rangle.\ \Box
\end{split}\end{equation*}

\begin{exam}Continuation of Example~\ref{ex2}. For the case of $p<1$, an
easy calculation yields:
$$(I-M)^{-2}=\left(\begin{array}{cccc}1 & 0 & 0 & 1+\frac{1}{1-p}\\ 0 & 1 & 0 & 0\\ 0& 0&
1& 0\\ 0& 0& 0& \frac{1}{(1-p)^{2}}\end{array}\right)$$ So, by
Proposition~\ref{prop} we see that the average running time of the
program considered in Examples~\ref{ex1} and~\ref{ex2} is
\begin{equation*}\begin{split}
\langle\Phi|N_0(I-M)^{2}(\rho_0\otimes I)|\Phi\rangle&=\langle
0|\rho_0\rangle +(1+\frac{1}{1-p})\langle 1|\rho_0|1\rangle\\
&=1+\frac{\langle 1|\rho_0|1\rangle}{1-p}.
\end{split}\end{equation*} In particular, if the initial state
$\rho_0=|\psi\rangle\langle\psi|$ with $|\psi\rangle=\alpha
|0\rangle+\beta |1\rangle$, then the average running time is
$1+\frac{|\beta|^{2}}{1-p}.$
\end{exam}

The techniques developed in this section can be further generalized
to analyze the long-run behaviors defined by de Alfaro~\cite{dA98}
and Br\'{a}zdil, Esparza and Ku\v{c}era~\cite{BEK05} for quantum
programs in finite-dimensional state spaces.

\section{Termination of Quantum Programs}\label{finite2}

It is usually not easy to check the termination condition (QV3) when
applying the quantum Sharir-Pnueli-Hart method. The purpose of this
section is to find some alternate conditions for termination of
quantum programs in finite-dimension state spaces employing the
ideas used in the last section.

As in the last section, we assume that the state space $H$ of
quantum programs considered in this section is finite-dimensional.

\begin{defn}Consider the program defined by quantum Markov chain
$(\mathcal{E},\rho_0)$ and termination test $\{M_0,M_1\}$.
\begin{enumerate}\item We say that the program terminates if the nontermination probability in the $n$th step
$$p^{N}_n=tr([\mathcal{E}_1\circ
(\mathcal{E}\circ\mathcal{E}_1)^{n-1}](\rho_0))=0$$ for some
positive integer $n$, where $\mathcal{E}_i$ $(i=0,1)$ are defined by
Eq.(\ref{super01}). \item We say that the program almost terminates
if $\lim_{n\rightarrow\infty}p^{N}_n=0,$ where $p^{N}_n$ is as in
item 1 for every $n\geq 1$ (also see Eq.~(\ref{prd2})).
\end{enumerate}
\end{defn}

It is obvious that almost termination is equivalent to (QV3) with
$Q=$ the identity operator $I$ on $H$. On the other hand, almost
termination implies (QV3) for every $Q$.

The above definition is a generalization of Definition 3.1
in~\cite{YF10}, where only the case that $\mathcal{E}$ is a unitary
operator, i.e. for some unitary operator $U$,
$\mathcal{E}(\rho)=U\rho U^{\dag}$ for all density operators $\rho$,
is considered.

The following lemma gives a simple termination condition in terms of
the matrix representation of super-operators.

\begin{lem}\label{term1}We consider the program defined by Markov chain $(\mathcal{E},\rho_0)$
and termination test $\{M_0,M_1\}$. Let $M$ be defined by
Eq.(\ref{mmatrix}), and let $|\Phi\rangle=\sum_j|jj\rangle$ be the
(unnormalized) maximally entangled state in $H\otimes H$. Then we
have:
\begin{enumerate}
\item The program terminates if and only if $M^{n}(\rho_0\otimes
I)|\Phi\rangle=0$ for some integer $n\geq 0$;\item The program
almost terminates if and only if
$$\lim_{n\rightarrow\infty}M^{n}(\rho_0\otimes I)|\Phi\rangle=0,$$
\end{enumerate}\end{lem}

\textit{Proof:} We only prove the first conclusion, the second is
similar. Since $\mathcal{E}$ is trace-preserving, it follows from
Eq.~(\ref{meq2}) that
\begin{equation*}\begin{split}
tr([\mathcal{E}_1\circ
(\mathcal{E}\circ\mathcal{E}_1)^{n-1}](\rho_0))&=tr(
(\mathcal{E}\circ\mathcal{E}_1)^{n}](\rho_0))\\
&=\langle\Phi|M^{n}(\rho_o\otimes I)|\Phi\rangle.
\end{split}\end{equation*} Moreover, it is clear that
$\langle\Phi|M^{n}(\rho_o\otimes I)|\Phi\rangle=0$ if and only if
$M^{n}(\rho_o\otimes I)|\Phi\rangle=0.\ \Box$

We now turn to consider the terminating problem of quantum programs.
Recall that a super-operator $\mathcal{E}$ together with a
termination test $\{M_0,M_1\}$ can be understood as a program
scheme.

\begin{defn} The program scheme is terminating (resp. almost terminating) if for
every density operator $\rho$, the program generated from the
program scheme with initial state $\rho$, i.e. the program defined
by Markov chain $(\mathcal{E},\rho)$ and termination test
$\{M_0,M_1\}$, terminates (resp. almost terminates).
\end{defn}

Similar to Lemma~\ref{term1}, we have:

\begin{lem}\label{term2}We consider the program scheme defined by super-operator
$\mathcal{E}$ and termination test $\{M_0,M_1\}$.
\begin{enumerate}\item The program scheme is terminating if and only
if $M^{n}|\Phi\rangle=0$ for some integer $n\geq 0$;\item The
program is almost terminating if
$\lim_{n\rightarrow\infty}M^{n}|\Phi\rangle=0,$\end{enumerate}where
$M$ and $|\Phi\rangle$ are as in Lemma~\ref{term1}.
\end{lem}

\textit{Proof:} Notice that a program scheme is terminating if and
only if it terminates when the initial state is $\rho_0=I/\dim H.$
Then this lemma follows immediately from Lemma~\ref{term1}. $\Box$

We now present the main result of this section.

\begin{prop}Consider the program scheme defined by super-operator
$\mathcal{E}$ and termination test $\{M_0,M_1\}$. Let $M$ and
$|\Phi\rangle$ be as in Lemma~\ref{term1}. Then we have:
\begin{enumerate}
\item For any integer $k\geq$ the maximal size of Jordan blocks of $M$ corresponding to eigenvalue $0$,
the program is terminating if and only if $M^{k}|\Phi\rangle=0.$
\item The program is almost terminating if and only if $|\Phi\rangle$ is
orthogonal to all the eigenvectors of $M$ corresponding to
eigenvalue with module $1.$
\end{enumerate}
\end{prop}

\textit{Proof:} For part 1, if $M^{k}|\Phi\rangle=0$, then by
Lemma~\ref{term2} we conclude that the program is terminating.
Conversely, suppose that the program is terminating. Again by
Lemma~\ref{term2}, there exists some integer $n\geq 0$ such that
$M^n|\Phi\rangle=0$. We want to show that $M^{k}|\Phi\rangle=0.$
Without any loss of generality, we assume the Jordan decomposition
$M=SJ(M)S^{-1}$, where
$$J(M)=diag(J_{k_1}(\lambda_1),J_{k_2}(\lambda_2),\cdot\cdot\cdot,J_{k_l}(\lambda_l))$$
with $|\lambda_1|\geq\cdot\cdot\cdot\geq|\lambda_s|>0,\
\lambda_{s+1}=\cdot\cdot\cdot=\lambda_l=0.$ Observe that
$M^n=SJ^n(M)S^{-1}.$ Since $S$ is nonsingular, it follows from
$M^n|\Phi\rangle=0$ that $J^n(M)S^{-1}|\Phi\rangle=0.$ We can write
\begin{eqnarray*}
J(M)=\left(
\begin{array}{cc}
 A & 0  \\
 0 & B \\
\end{array}
\right),\ \ S^{-1}|\Phi\rangle=\left(
\begin{array}{cc}
 |x\rangle   \\
 |y\rangle \\
\end{array}
\right),
\end{eqnarray*}
where $$A=diag(J_{k_1}(\lambda_1),...,J_{k_s}(\lambda_s)),$$
$$B=diag(J_{k_{s+1}}(0),...,J_{k_l}(0)),$$ $|x\rangle$ is a
$t-$dimensional vector, $|y\rangle$ is a $(d^2-t)-$dimensional
vector, and $t=\sum_{j=1}^sk_j$. Then
\begin{eqnarray*}
J(M)^nS^{-1}|\Phi\rangle=\left(
\begin{array}{cc}
 A^n|x\rangle   \\
 B^n|y\rangle  \\
\end{array}
\right).
\end{eqnarray*}Note that $\lambda_1,...,\lambda_s\neq 0$. So,
$J_{k_1}(\lambda_1),...,J_{k_s}(\lambda_s)$ are nonsingular, and $A$
is nonsingular too. Thus, $$J^n(M)S^{-1}|\Phi\rangle=0$$ implies
$A^n|x\rangle=0$ and furthermore $|x\rangle=0$. On the other hand,
for each $j$ with $s+1\leq j\leq l$, since $k\geq k_j$, it holds
that $J_{k_j}^k(0)=0$. Consequently, $B^k=0$,
$J^k(M)S^{-1}|\Phi\rangle=0$, and
$$M^k|\Phi\rangle=SJ^k(M)S^{-1}|\Phi\rangle=0.$$

For part 2, by Lemma~\ref{term2}, the program is almost terminating
if and only if
$$\lim\limits_{n\to\infty}J^n(M)S^{-1}|\Phi\rangle=0.$$ We assume
that
$$1=|\lambda_1|=\cdot\cdot\cdot=|\lambda_r|>|\lambda_{r+1}|\geq\cdot\cdot\cdot\geq|\lambda_l|$$
in the Jordan decomposition of $M$, and we write:
\begin{eqnarray*}
J(M)=\left(
\begin{array}{cc}
 C & 0  \\
 0 & D \\
\end{array}
\right),\ \ S^{-1}|\Phi\rangle=\left(
\begin{array}{cc}
 |u\rangle   \\
 |v\rangle  \\
\end{array}
\right)
\end{eqnarray*}
where
\begin{equation*}\begin{split}C&=diag(\lambda_1,...,\lambda_r),\\
D&=diag(J_{k_{r+1}}(\lambda_{r+1}),...,J_{k_l}(\lambda_{l})),\end{split}\end{equation*}
$|u\rangle$ is an $r-$dimensional vector, and $|v\rangle$ is a
$(d^2-r)-$dimensional vector. (Note that
$J_{k_1}(\lambda_1),...,J_{k_r}(\lambda_r)$ are all $1\times 1$
matrices because $|\lambda_1|=...=|\lambda_r|=1$; see
Lemma~\ref{tech0}.)

If $|\Phi\rangle$ is orthogonal to all the eigenvectors of $M$
corresponding to eigenvalue with module $1,$ then $|u\rangle=0$. On
the other hand, for each $j$ with $r+1\leq j\leq l$, since
$|\lambda_j|<1$, we have
$$\lim_{n\rightarrow\infty}J_{k_j}^n(\lambda_j)=0.$$ Thus,
$\lim\limits_{n\to\infty}D^n=0.$ So, it follows that
\begin{eqnarray*}
\lim\limits_{n\to\infty}J(M)^nS^{-1}|\Phi\rangle=\lim\limits_{n\to\infty}\left(
\begin{array}{cc}
 C^n|u\rangle   \\
 D^n|v\rangle  \\
\end{array}
\right)=0.
\end{eqnarray*}
Conversely, if
$$\lim\limits_{n\to\infty}J(M)^nS^{-1}|\Phi\rangle=0,$$ then
$\lim_{n\rightarrow\infty}C^n|u\rangle=0$. This implies
$|u\rangle=0$ because $C$ is a diagonal unitary. Consequently,
$|\Phi\rangle$ is orthogonal to all the eigenvectors of $M$
corresponding to eigenvalue with module $1.$ $\Box$

The above two propositions considerably generalize Corollaries 5.2
and 6.2 in~\cite{YF10}.

\section{Conclusion}\label{concl}

This paper develops a methodology for verifying quantum programs in
which quantum programs are modeled by quantum Markov chains and
verified properties are described in terms of Hermitian operators.
The Sharir-Pnueli-Hart method for verifying probabilistic programs
is generalized into the quantum setting. For quantum programs with
finite-dimensional state spaces, a simpler verification method is
found, which is especially useful for analysis of long-run behavior
like the average running time. The methodology developed in this
paper mainly targets verification of programs for future quantum
computers, but it also has potential applications in other areas
such as verifications of quantum communication protocols and
engineered quantum systems~\ref{DM03}. Further studies will
naturally go along two lines:\begin{itemize}
\item \textit{More Complex Programs}: An interesting
topic in this direction is verification of quantum programs with
recursive procedures. A possible model of these programs is a
quantum generalization of recursive state machines introduced by
Alur et al.~\cite{Al05} and recursive Markov chains introduced by
Etessami and Yannakakis~\cite{EY09}.

\item \textit{More Sophisticated
Properties}: To express properties of quantum programs more
sophisticated than those considered in this paper, a possible way is
to define quantum extension of temporal logics. Baltazar et
al.~\cite{BC08} already proposed a quantum computation tree logic,
but more research in this direction is in order because some
fundamental problems are still not well-understood, e.g. how can the
notions of sequential and joint measurements that have puzzled
physicists for many years be incorporated into temporal logic?
\end{itemize}

\appendix
\section{Proof of Technical Lemmas}

\subsection{Proof of Lemma~\ref{tech0}}

Recall that the quantum program is defined by Markov chain
$(\mathcal{E},\rho_0)$ together with termination test $\{M_0,M_1\}$,
and the operator-sum representation of super-operator $\mathcal{E}$
is as follows: $\mathcal{E}(\rho)=\sum_iE_i\rho E_i^{\dag}$ for all
partial density operators $\rho$. Assume that the dimension of the
state space $H$ is $d=\dim H$. We write $\mathcal{G}
=\mathcal{E}\circ\mathcal{E}_1$ and
$\mathcal{E}_i(\rho)=M_i\rho{M_i}^{\dag}$ for any partial density
operator $\rho$ and $i=0$, $1$.

\begin{lem}The super-operator $\mathcal{G}+\mathcal{E}_0$ is trace-preserving:
\begin{equation}\label{tr-p}tr[(\mathcal{G}+\mathcal{E}_0)(\rho)]=tr(\rho)\end{equation} for all
partial density operators.\end{lem}

\textit{Proof:} It suffices to see that
\begin{equation*}\begin{split}\sum_i(E_iM_1)^{\dag}&E_iM_1+{M_0}^{\dag}M_0\\ &={M_1}^{\dag}
(\sum_i{E_i}^{\dag}E_i)M_1+{M_0}^{\dag}M_0\\
&={M_1}^{\dag}M_1+{M_0}^{\dag}M_0=I.\ \Box\end{split}\end{equation*}

\begin{lem}\label{m-dec}For any matrix $A$, there are positive matrices
$B_1,$ $B_2,B_3,B_4$ such that \begin{enumerate}\item
$A=B_1-B_2+iB_3-iB_4$; and \item $trB_i^2\leq tr(A^{\dag}A)$
$(i=1,2,3,4).$
\end{enumerate}
\end{lem}

\textit{Proof:} We can take Hermitian operators
$$(A+A^{\dag})/2=B_1-B_2,\ \
-i(A-A^{\dag})/2=B_3-B_4,$$ where $B_1,B_2$ are positive operators
with orthogonal supports, and $B_3,B_4$ are positive operators with
orthogonal supports. Then it holds that
\begin{equation*}\begin{split}\sqrt{\mathrm{tr}{B_1}^{2}}&=\sqrt{\mathrm{tr}({B_1}^{\dag}B_1)}\\
&\leq\sqrt{\mathrm{tr}({B_1}^{\dag}B_1+{B_2}^{\dag}B_2)}\\
&=\|((A+A^{\dag})/2\otimes I) |\Phi\rangle \|\\
&\leq (\|(A\otimes I)|\Phi\rangle \|+\|(A^\dag\otimes I)|\Phi\rangle\|)/2\\
&=\sqrt{\mathrm{tr}({A}^{\dag}A)}.\end{split}
\end{equation*} It is similar to prove that $\mathrm{tr}{B_i}^{2}\leq \mathrm{tr}(A^{\dag}A)$ for
$i=2,3,4$. $\Box$

\begin{lem}\label{A1}For any integer $n\geq 0$, and for any $|\alpha\rangle$ in $H\otimes H$, we have:
$$\|M^n |\alpha\rangle\|\leq 4\sqrt{d}\| |\alpha\rangle\|.$$\end{lem}

\textit{Proof:} Suppose that
$|\alpha\rangle=\sum_{i,j}a_{ij}|ij\rangle.$ Then we can write:
$|\alpha\rangle=(A\otimes I)|\Phi\rangle,$ where $A=(a_{ij})$ is a
$d\times d$ matrix. A routine calculation yields: $\|
|\alpha\rangle\|=\sqrt{\mathrm{tr}A^{\dag}A}.$

We write: $A=B_1-B_2+iB_3-iB_4$ according to Lemma~\ref{m-dec}. The
idea behind this decomposition is that the trace-preserving property
Eq.~(\ref{tr-p}) only applies to positive operators. Put
$|\beta_i\rangle=(B_i\otimes I)|\Phi\rangle$ for $i=1,2,3,4$. Using
the triangle inequality, we obtain:
\begin{eqnarray*}
\|M^n|\alpha\rangle\|\leq
\sum_{i=1}^4\|M^n|\beta_i\rangle\|=\sum_{i=1}^4\|(\mathcal{F}^m(B_i)\otimes
I)|\Phi\rangle\|.
\end{eqnarray*}
Note that \begin{equation}\label{pp1}\|(\mathcal{F}^m(B_i)\otimes
I)|\Phi\rangle\|=\sqrt{\mathrm{tr}(\mathcal{F}^m(B_i))^2},\end{equation}
\begin{equation}\label{pp2}\mathrm{tr}\rho^2\leq (\mathrm{tr}\rho)^2,\end{equation}
\begin{equation}\label{pp3}\mathrm{tr}[\mathcal{F}^m(\rho)]\leq
\mathrm{tr}[(\mathcal{F}+\mathcal{E}_0)^m(\rho)]=\mathrm{tr}\rho.\end{equation}
Combining Eqs.~(\ref{pp1}), (\ref{pp2}) and (\ref{pp3}), one would
see that
\begin{eqnarray*} \sqrt{\mathrm{tr}(\mathcal{F}^m(B_i))^2}\leq
\sqrt{ (\mathrm{tr}\mathcal{F}^m(B_i))^2}\leq
\sqrt{(\mathrm{tr}B_i)^2}.
\end{eqnarray*}Furthermore,
by the Cauchy-inequality we have $(\mathrm{tr}\rho)^2\leq
d\cdot(\mathrm{tr}\rho^2).$ Therefore,
\begin{eqnarray*}
\|F^m|\alpha\rangle\|\leq\sum_{i=1}^4\sqrt{d\cdot\mathrm{tr}B_i^2}\leq
4 \sqrt{d\cdot\mathrm{tr}(A^{\dag}A)}=4\sqrt{d}\| |\alpha\rangle\|.
\Box
\end{eqnarray*}

Now we are ready to prove Lemma~\ref{tech0}. We prove the first part
by refutation. If there is some eigenvalue $\lambda$ of $M$ with
$|\lambda|>1$, suppose the corresponding normalized eigenvector is
$|x\rangle$: $M|x\rangle=\lambda |x\rangle.$ Choose integer $n$ such
that $|\lambda|^n>4\sqrt{d}.$ Then
$\|M^n|x\rangle\|=\|\lambda^n|x\rangle\|=|\lambda|^m>4\sqrt{d}\||x\rangle\|.$
This contradicts to Lemma~\ref{A1}.

The second part can also be proved by refutation. Without any loss
of generality, we assume that $|\lambda_1|=1$ with $k_1>1$ in the
Jordan decomposition of $M$. Suppose that
$\{|i\rangle\}_{i=1}^{d^2}$ is the orthonormal basis of $H\otimes H$
compatible with the numbering of the columns and rows of $M$. Take
an unnormalized vector $|y\rangle=S|k_1\rangle.$ Since $S$ is
nonsingular, there are real numbers $R,r>0$ such that $r\cdot
\||x\rangle\|\leq\|S|x\rangle\|\leq R\cdot \| |x\rangle\|$ for any
vector $|x\rangle$ in $H\otimes H$. By definition, it holds that $\|
|y\rangle\|\leq R.$ We can choose integer $n$ such that $nr>R\cdot
4\sqrt{d}$ because $r>0$. Then a routine calculation yields:
\begin{eqnarray*}
M^m |y\rangle=S\sum_{t=1}^{k_1}\left(\begin{array}{cc}n\\
t\end{array}\right){\lambda_1}^{n-t}|k_1-t\rangle,
\end{eqnarray*}
Consequently, we have: \begin{equation*}\begin{split}\|M^n
|y\rangle\|&\geq
r\cdot\sum_{t=1}^{k_1}\left(\begin{array}{cc}n\\
t\end{array}\right)|\lambda_1|^{n-t}\geq nr\\ & >R\cdot
4\sqrt{d}\geq 4\sqrt{d}\| |y\rangle\|.\end{split}\end{equation*}
This contradicts to Lemma~\ref{A1} again.

\subsection{Proof of Lemma~\ref{tech}}
Without any loss of generality, we assume
$1=|\lambda_1|=\cdot\cdot\cdot=|\lambda_s|>|\lambda_{s+1}|\geq
\cdot\cdot\cdot\geq|\lambda_l|.$ Then
\begin{eqnarray*}
J(M)=\left(
\begin{array}{cc}
 U & 0  \\
 0 & J_1 \\
\end{array}
\right),
\end{eqnarray*}
where $U=diag(\lambda_1,\cdot\cdot\cdot,\lambda_s)$ is an $s\times
s$ diagonal unitary, and
$$J_1=\mathrm{diag}(J_{k_{s+1}}(\lambda_{s+1}),\cdot\cdot\cdot,J_{k_l}(\lambda_l)).$$
Moreover, we have:
\begin{eqnarray*}
J(N)=\left(
\begin{array}{cc}
 0 & 0  \\
 0 & J_1 \\
\end{array}
\right).
\end{eqnarray*}

The convergence of
$\sum_{n=0}^\infty(\mathcal{E}_0\circ\mathcal{G}^n)$ follows
immediately from Proposition~\ref{conve}, and it in turn implies the
convergence of $\sum_{n=0}^\infty N_0M^n.$ It is clear that
\begin{eqnarray*}
\sum_{n=0}^\infty N_0M^n=\sum_{n=0}^\infty N_0SJ(M)^nS^{-1}.
\end{eqnarray*} Since $S$ is nonsingular,
we see that $\sum_{n=0}^\infty N_0SJ(M)^n$ converges. This implies
that $\lim\limits_{n\to\infty} N_0SJ(M)^n=0.$ Now we write:
\begin{equation*}
N_0S=\left(
\begin{array}{cc}
 Q & R  \\
 S & T \\
\end{array}
\right),\end{equation*} where $Q$ is an $s\times s$ matrix, $T$ is a
$(d^2-s)\times (d^2-s)$ matrix, and $d=\dim H$ is the dimension of
the state space $H$. Then
\begin{equation*} N_0SJ(M)^n=\left(
\begin{array}{cc}
 QU^n & R{J_1}^n  \\
 SU^n & T{J_1}^n \\
\end{array}
\right),
\end{equation*} and it follows that
$\lim_{n\rightarrow\infty}QU^n=0$ and
$\lim_{n\rightarrow\infty}SU^n=0.$ So, we have: $tr(Q^\dag
Q)=\lim\limits_{n\to\infty} \mathrm{tr} (QU^n)^{\dag}QU^n=0,$ and
$tr(S^\dag S)=\lim\limits_{n\to\infty}
\mathrm{tr}(SU^n)^{\dag}SU^n=0.$ This yields $Q=0$ and $S=0$, and it
follows immediately that $N_0 M^n=N_0N^n$.

\bigskip\

\textit{Acknowledgment:} This work was partly supported the
Australian Research Council (Grant No: DP110103473) and the National
Natural Science Foundation of China (Grant No: 60736011).

\end{document}